\documentclass[11pt]{article}

\usepackage[final]{acl}

\usepackage{times}
\usepackage{latexsym}
\usepackage[table]{xcolor}  
\usepackage[T1]{fontenc}
\usepackage{amsfonts}
\usepackage{amsmath}
\usepackage{booktabs}
\usepackage{multirow}
\usepackage{float}

\usepackage[utf8]{inputenc}

\usepackage{microtype}

\usepackage{inconsolata}

\usepackage{graphicx}
\usepackage[percent]{overpic}

\usepackage{xcolor}

\author{
 \textbf{Yixu Huang\textsuperscript{\rm $\alpha$}\textsuperscript{\dag}},
 \textbf{Bo Li\textsuperscript{\rm $\alpha$}\textsuperscript{\dag}},
 \textbf{Na Li\textsuperscript{\rm $\beta$}\textsuperscript{\dag}},
 \textbf{Zhe Wang\textsuperscript{\rm $\beta$}},
 \textbf{Kaijie Chen\textsuperscript{\rm $\gamma$}},
 \textbf{Haonan Ge\textsuperscript{\rm $\delta$}},
\\
 \textbf{Qingyi Si\textsuperscript{\rm $\eta$}},
 \textbf{Yuanzhe Shen\textsuperscript{\rm $\alpha$}},
 \textbf{Ruihan Yang\textsuperscript{\rm $\alpha$}},
 \textbf{Guangjing Wang\textsuperscript{\rm $\beta$}\textsuperscript{\ddag}},
 \textbf{Hongcheng Guo\textsuperscript{\rm $\alpha$}\textsuperscript{\ddag}}
\\
\\
 \textsuperscript{\rm $\alpha$}Fudan University,
 \textsuperscript{\rm $\beta$}Xiaohongshu Inc.,
 \textsuperscript{\rm $\gamma$}Tongji University,
\\
 \textsuperscript{\rm $\delta$}University of California, Santa Barbara,
 \textsuperscript{\rm $\eta$}JD.COM
\\
 \small{
   \textsuperscript{\dag}Equal contribution \quad
   \textsuperscript{\ddag}Corresponding authors \quad
   \textbf{Correspondence:} \href{mailto:yixuhuang23@m.fudan.edu.cn,21302010068@m.fudan.edu.cn}{\{yixuhuang23, lib21\}@m.fudan.edu.cn}
 }
}

\begin{document}

\title{
GUI Agents for Continual Game Generation
}

\maketitle

\begin{abstract}
Generating a game is not the same as making one playable. Existing code-generation approaches often translate a prompt directly into an artifact, leaving interaction-level failures undetected. We argue that game generation requires a player and study two roles for graphical user interface (GUI) agents. First, we introduce \textbf{PlaytestArena}, an evaluation environment containing 200 browser-based game-generation tasks across eight genres, each paired with rubrics of expected in-play behaviors. An independent GUI judge loads and plays each build to adjudicate these rubrics. Second, we propose \textbf{Play2Code}, in which a game agent and a rubric-blind GUI playtester iteratively generate, play, and refine games through shared memory. The playtester provides gameplay traces and actionable feedback, while a separate GPT-5.5 judge assigns final benchmark scores. Across three frontier backbones, Play2Code achieves a 66.8\% rubric pass rate, outperforming single-pass and agentic-coding baselines by 37.1 and 14.6 points, respectively. Its scores also improve monotonically across refinement rounds. Further analysis shows that GUI-agent feedback is fully logged and traceable, while its priorities vary substantially across model backbones. These results establish GUI playtesting as an evaluation and refinement signal for interactive code generation. Our project website is available at \url{https://continual-game-generation.vercel.app/}.
\end{abstract}

\section{Introduction}

\begin{center}
\emph{``A game is a series of interesting choices.''}
\end{center}
\hfill---\,Sid Meier

\vspace{1em}

A musical score, by itself, is silent. Every note in place, every dynamic marked, every tempo specified. Yet whether the music is moving, dull, or merely correct cannot be known until someone plays it. A game, like a score, must be played. It can compile, run, and pass every test, yet be broken in ways no static analysis can reveal. This is why the game industry has long relied on human playtesters, people who sit down with a build, play for hours, and report back what no inspection can see, such as a button press that gets no response, a sprite that fails to render, or a win condition that never triggers.

Recent LLM-based code generators have begun to produce increasingly capable programs from natural-language prompts~\citep{zheng2023codegeex,zhu2024deepseek,dong2025survey}, including interactive web game artifacts~\citep{hu2024game,gallotta2024large,sweetser2024large}. Yet this progress has outpaced playtesting itself: generators produce candidate games faster than any human can play them, while what makes a game good remains as inaccessible to inspection as it ever was. In most code generation tasks, quality can be checked against tests or specifications. In game generation, it cannot. Under this constraint, how to evaluate and improve game generation remains an open question~\citep{hu2024game}.

\begin{figure*}[t]
    \centering
    \includegraphics[width=\textwidth]{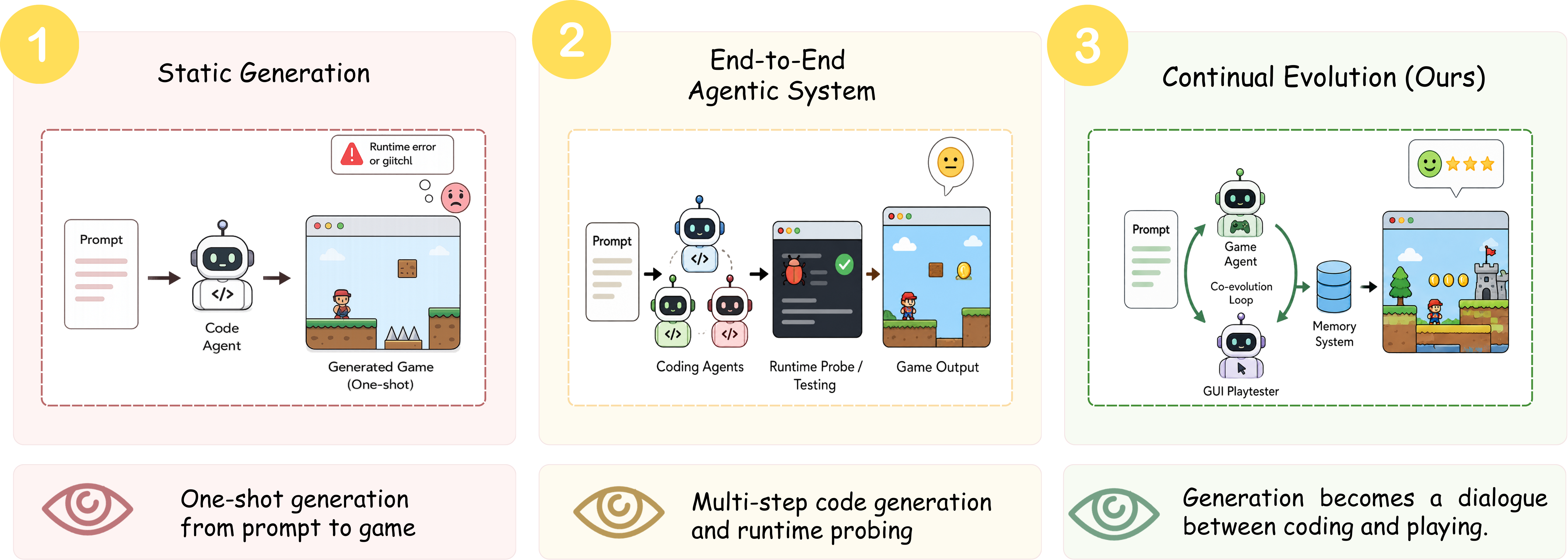}
\caption{Three paradigms for AI-driven game generation. Unlike prior one-shot or runtime-probing pipelines, Play2Code makes generation a sustained dialogue between coding and playing.}
\label{fig:three_paradigms}
\vspace{-1em}
\end{figure*}

We argue that evaluating and refining game generation requires a player. Among the candidate surrogates one might imagine, only GUI agents meet the game on the same presentational surface a human playtester does: they perceive the rendered interface, decide on actions grounded in that perception, and act through the same input channels a human would. We therefore study two roles for GUI agents in game generation. 

First, as an automated rubric evaluator, we introduce PlaytestArena, a new environment for studying game generation when quality lives in play rather than code (§3). Our arena consists of a curated set of generation tasks, each pairing a natural-language prompt with a rubric of expected in-play behaviors that defines what the game should look like when correctly implemented. An independent GUI judge loads each generated game in a browser, plays it through clicks and keys, and adjudicates each rubric criterion based on what it observes during play. The tasks span common game genres, including puzzle, strategy, and card games, each challenging generators with different demands on stateful gameplay, input handling, and visual feedback. PlaytestArena is useful for studying how generation methods perform when the unit of evaluation shifts from code to play.

Second, as a \textbf{refinement playtester}, we introduce \textbf{Play2Code}, a system that bridges game generation and playtesting, and sustains their interaction to refine the game (§\ref{sec:play2code}). As specified in Figure~\ref{fig:three_paradigms}, prior work spans direct synthesis and agentic generation pipelines~\citep{hu2024game,jiang2026opengame}. Related solver--verifier systems refine code through iterative execution feedback~\citep{wang2026co,huang2026ace}; Play2Code instead uses what the current build is like to play. A Game Agent modifies the code, while a GUI playtester interacts with the rendered build and returns only gameplay traces, observations, and actionable fixes. The playtester never receives the PlaytestArena rubric or produces the final benchmark score; scoring is performed afterward by the independent GUI judge. Generation thus becomes not a one-shot translation but a sustained dialogue between coding and playing.

In this paper, we study the following research questions:

\textbf{RQ1}: How does the nature of games reshape what counts as ``better''?

\textbf{RQ2}: What must a ``player'' look like, and where should it sit?

\textbf{RQ3}: How does such a ``player'' drive game generation, and does it drive it where a human would?

To study these questions, we make three complementary contributions. \textbf{First}, we build PlaytestArena, a testbed of 200 browser-based game generation tasks spanning eight genres, each paired with a rubric of expected in-play behaviors and adjudicated by an independent GPT-5.5 GUI judge. \textbf{On top of this}, we develop Play2Code, in which a Game Agent and a rubric-blind GUI playtester operate in a sustained loop with shared memory, letting experience accumulate across rounds and tasks. Across three frontier backbones, Play2Code achieves a 66.8\% rubric pass-rate, improving over single-pass and agentic-coding baselines by 37.1 and 14.6 points respectively, with rubric scores rising monotonically across rounds. \textbf{Lastly}, we find that the GUI playtester's feedback is fully logged and traceable, while its priorities vary substantially across model backbones.

\section{Can GUI Agents Playtest?}
\label{sec:gui_playtest}

\begin{figure*}[t]
    \centering
    \includegraphics[width=\textwidth]{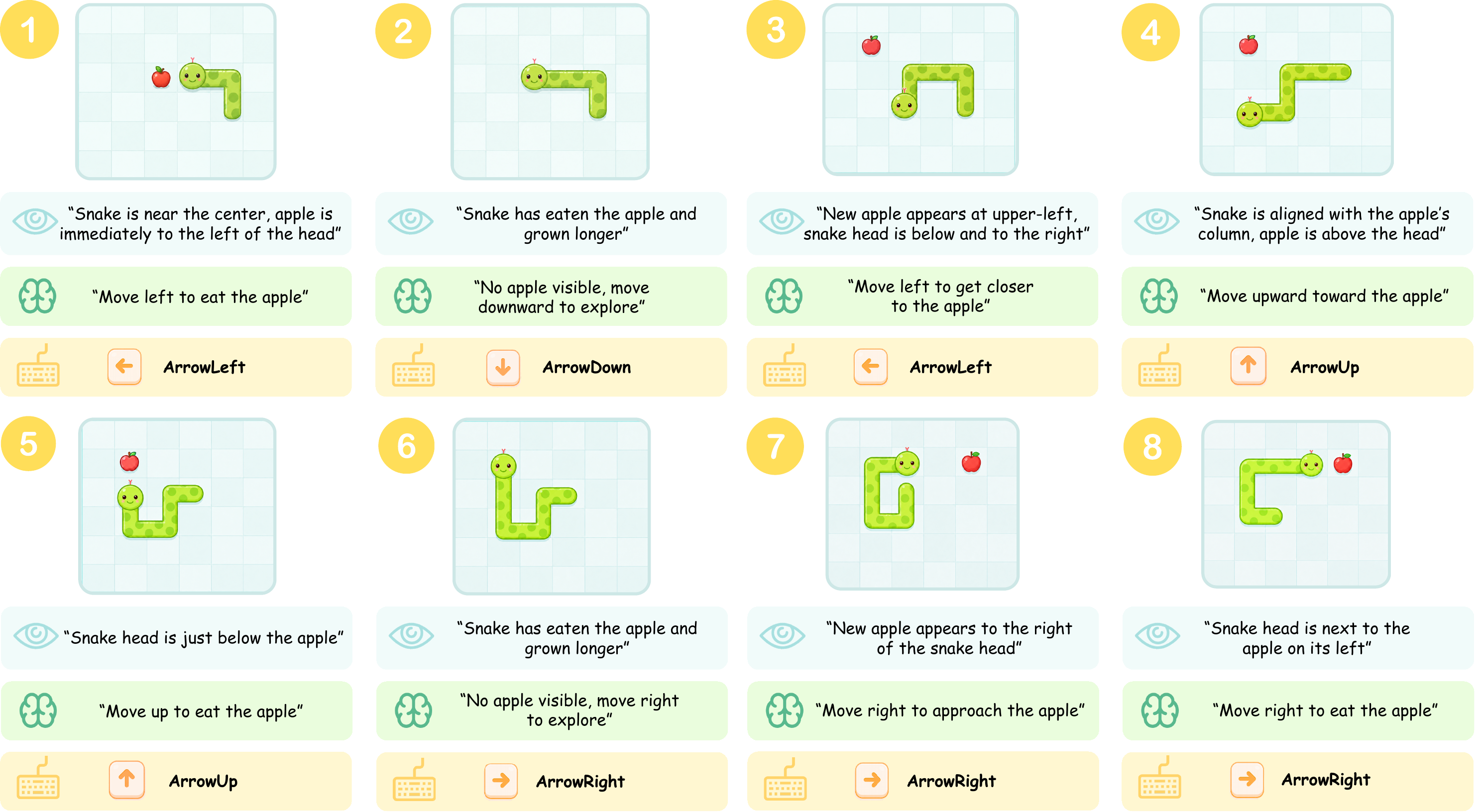}
    \caption{An example GUI agent session playing Snake. At each grid step, the agent observes the current screen, reasons about the next move, and issues a keyboard input. Across eight steps, the agent tracks snake and apple positions, updating its plan after each state transition.
    }
    \label{fig:gui_play}
    \vspace{-1em}
\end{figure*}

Consider how a human playtester works. They open the build, sit at the screen, and engage with the game in real time. Their access is mediated entirely by its presentational surface. What they produce is a record of what the game was like to play, including which mechanics worked, where the experience broke down, and what felt missing.

This is also where GUI agents live. A GUI agent perceives the rendered interface, decides on actions grounded in that perception, and acts through the same input channels a human would: clicks, key presses, scrolls. Among the candidate surrogates one might imagine, execution-free code critics~\citep{yadavally2025large}, environment-specific automated playtesters~\citep{shin2020playtesting,balla2024pytag}, or GUI agents themselves~\citep{wang2024gui,ouyang2026gameworld}, only the last meets the game on the same general-purpose visual and input surface where human playtesting takes place. We therefore take GUI agents as the natural surrogate for human playtesters.

Surface compatibility, however, is only a necessary condition. Whether GUI agents can actually playtest remains to be shown. We verify this directly by assembling a small testbed of 20 browser-based games, curated from both LLM-generated games and public web games, spanning puzzle, strategy, card, platformer, and management games. Because games vary widely in length and structure, each game is decomposed into several discrete levels, yielding approximately 120 levels in total. An agent is considered to have passed a level if it satisfies the level’s completion condition within an episode. We connect three GUI agent backbones (GPT-5.4~\citep{openai2026gpt54}, Claude Sonnet 4.6~\citep{anthropic2026claude}, and Kimi K2.5~\citep{moonshot2026kimi}) to each game and let them play independently. As a reference point, we recruit three human players to play the same levels under matched conditions. See more details in Appendix \S\ref{app:gui_playtest_feasibility}.

Table~\ref{tab:playtest_passk} reports pass@k for each backbone and the human reference at $k \in \{5, 10, 20\}$. All three GUI agents clear the majority of levels. GPT-5.4 is the strongest evaluated GUI agent, while remaining 10 percentage points below humans at pass@20. Beyond the aggregate numbers, GUI agents exhibit the three capabilities a playtester needs: (1) \textbf{They observe}: their state descriptions reflect the rendered game screen, including object locations, interface changes, and failure states. (2) \textbf{They reason}: their intermediate decisions connect the current observation to the next intended goal, such as approaching a target, avoiding an obstacle, or retrying after a failed interaction. (3) \textbf{They act}: their keyboard and mouse inputs follow from this reasoning and produce observable state transitions in the game. Figure~\ref{fig:gui_play} walks through one such session, showing the agent's observations, reasoning, and actions over a playing trajectory. This illustrates that the agent is not merely judging a static artifact, but genuinely interacting with the game and, when needed, producing signals for evaluation and refinement based on the trajectory.

\begin{table}[h]
\centering
\small
\caption{Pass@$k$ for GUI agent backbones and the human reference.}
\label{tab:playtest_passk}
\begin{tabular}{@{}lccc@{}}
\toprule
\textbf{Agent} & \textbf{pass@5} & \textbf{pass@10} & \textbf{pass@20} \\
\midrule
GPT-5.4            & 0.67 & 0.75 & 0.82 \\
Sonnet 4.6  & 0.63 & 0.76 & 0.79 \\
Kimi K2.5          & 0.52 & 0.68 & 0.72 \\
\midrule
Human Reference    & 0.83 & 0.92 & 0.92 \\
\bottomrule
\end{tabular}
\vspace{-1em}
\end{table}

\begin{figure}[t]
    \centering
    \includegraphics[width=\linewidth]{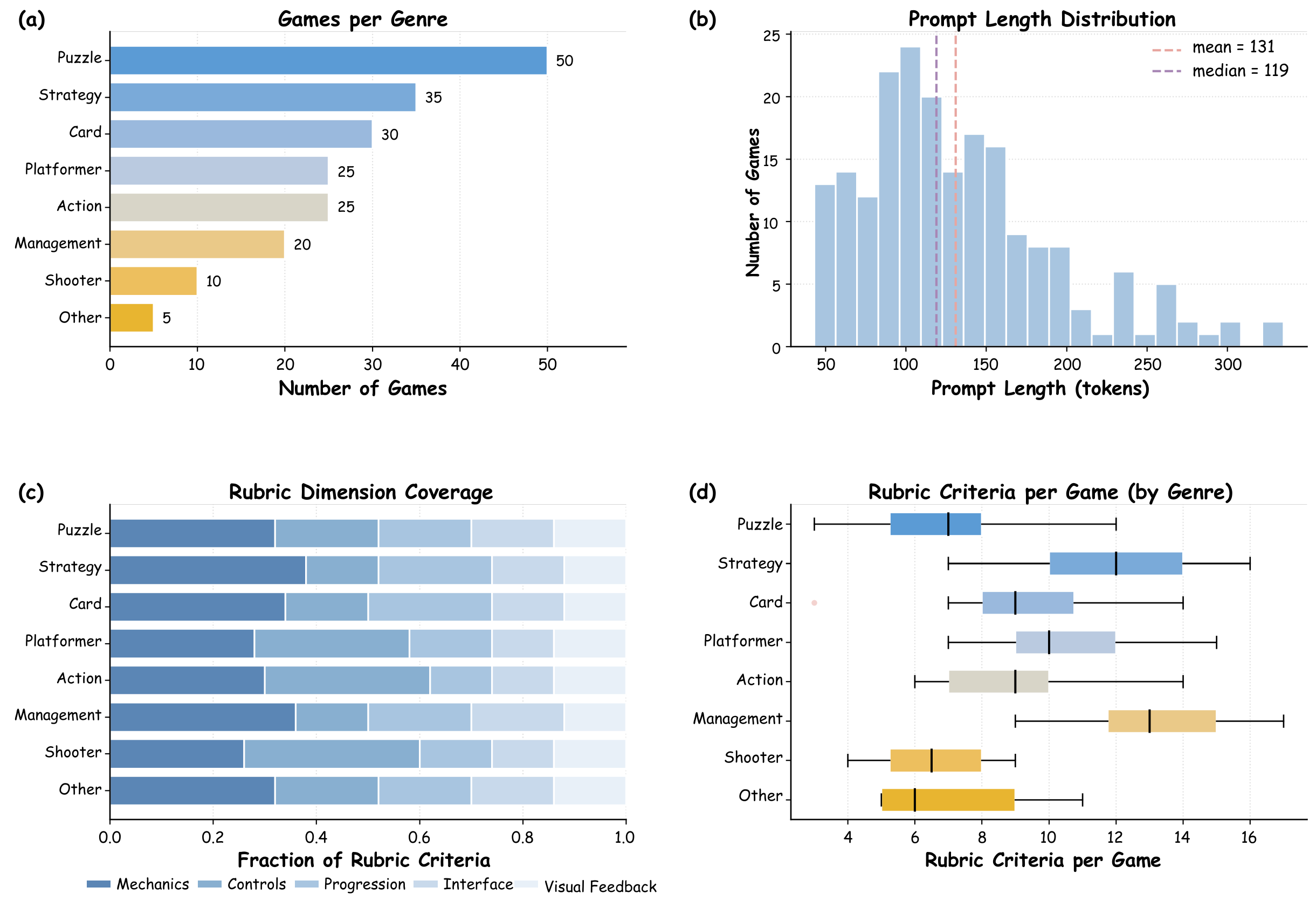}
   \caption{Statistics of PlaytestArena: (a) distribution of 200 games across eight genres, (b) prompt length distribution, (c) per-genre coverage of five rubric dimensions, and (d) rubric criterion count per game by genre.}
    \label{fig:playtestarena_stats}
    \vspace{-1em}
\end{figure}

\section{What's “Better”: PlaytestArena}
\label{sec:playtestarena}
\begin{figure*}[t]
    \centering
    \includegraphics[width=\textwidth]{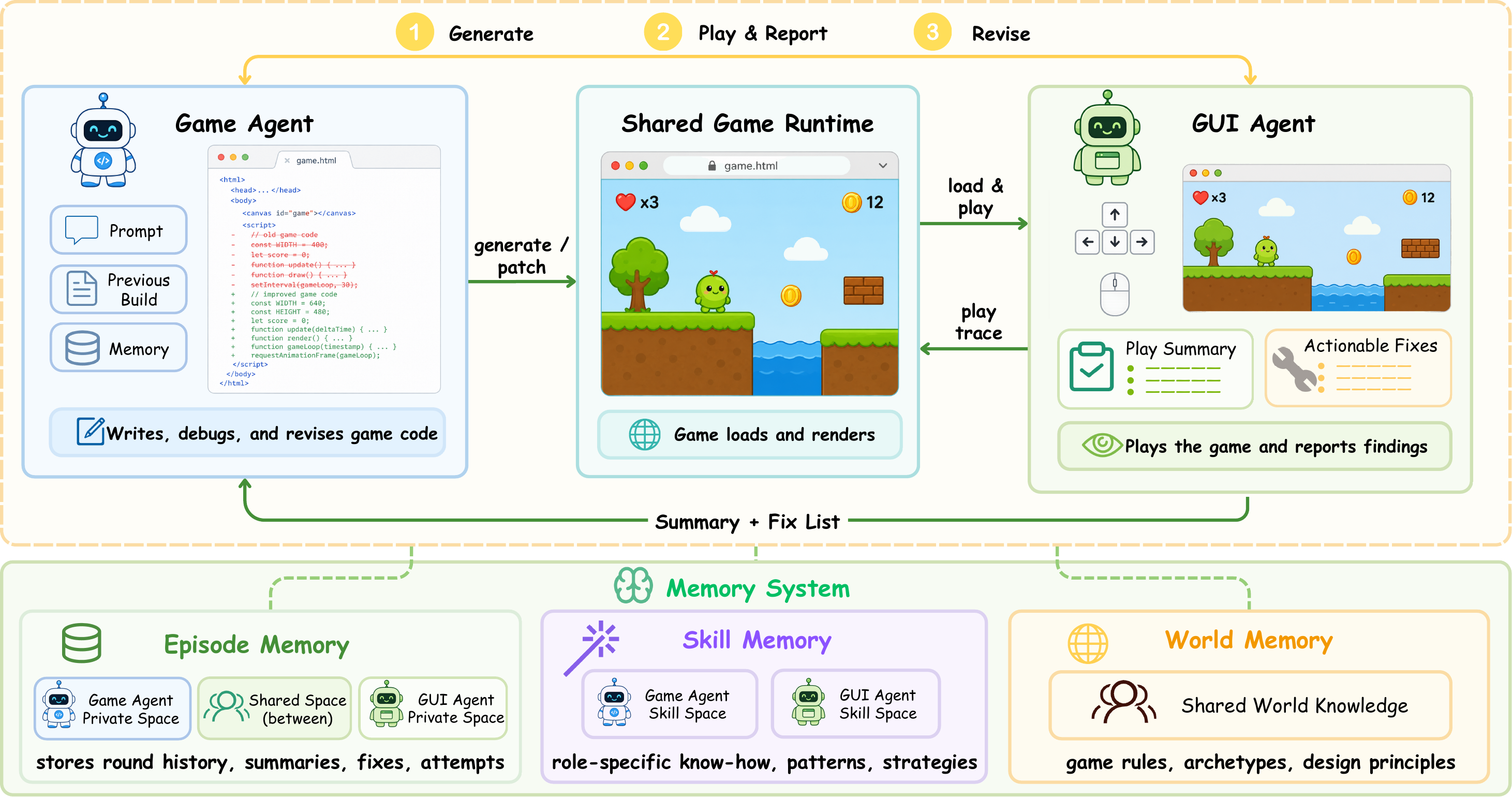}
    \caption{Overview of Play2Code. A game agent and GUI agent operate in a loop: the game agent generates or patches an HTML game, the GUI agent plays the build and produces a play summary and fix list, which guide the next revision. A three-layer memory system (episode, skill, and world memory) accumulates experience across rounds and tasks, turning GUI playtesting into a continual refinement signal.}
    \label{fig:play2code}
    \vspace{-1em}
\end{figure*}

Having established that GUI agents can carry out playtesting (\S\ref{sec:gui_playtest}), we now build the environment where they can judge game generation quality. Existing benchmarks for AI-generated games remain scarce. OpenGame-Bench~\citep{jiang2026opengame} and WebGameBench~\citep{zhang2026webgamebench}, for example, evaluate related aspects of browser-game generation but do not provide the same prompt-paired, criterion-level behavioral rubrics used here. PlaytestArena targets this complementary setting. Games are instantiated as self-contained HTML/CSS/JS bundles that run in any modern browser, placing game playtesting in the same substrate as the web GUIs that contemporary GUI agents already operate on. PlaytestArena requires two components: a set of target games, each paired with a generation prompt and a rubric defining what correct play should look like; and a playtester that inspects whether a generated game satisfies its rubric by playing it. The first defines what ``better'' means; the second turns that definition into a score.

\paragraph{Game Prompts Construction.} Each entry in PlaytestArena consists of a generation prompt and a corresponding rubric. Game topics are  collected from common game genres~\citep{wiki2026videogamegenres} and public web-game platforms~\citep{itchio2026,poki2026}, covering puzzle games, strategy games, card games, and others (Figure~\ref{fig:playtestarena_stats}a), totaling 200. For each topic, we recruited human experts, as described in Appendix \S\ref{app:playtestarena_construction}, to author prompts sketching the intended game, specifying stateful gameplay, win and loss conditions, player input, and feedback behavior. Prompts have a mean length of 131 tokens (median 119; Figure~\ref{fig:playtestarena_stats}b), concise enough to remain self-contained while specifying gameplay loops, win/loss conditions, and feedback behavior. Vague or non-self-contained prompts are removed in a final manual pass.

\paragraph{Rubrics Construction.} To build rubrics paired with prompts, we follow recent rubric-based evaluation practice~\citep{arora2025healthbench,li2026ueval}: human experts write criteria covering the intended mechanics, controls, progression, and visual feedback; they then refine these into testable items, each stating an expected behavior such as "the spawned enemies move toward the player's position". A final human pass ensures every item is observable and faithful to the prompt. The resulting rubric corpus contains 1{,}548 criteria across the 200 games, with a mean of 7.7 and a range of 5--17 criteria per game; coverage across the five dimensions is balanced across all eight genres, with Mechanics and Controls accounting for the largest share, consistent with the prompt-level emphasis on stateful gameplay  (Figure~\ref{fig:playtestarena_stats}c,d). The result is a per-game specification of what playing the generated artifact should reveal, against which any generation attempt can be checked.

\paragraph{Evaluation.} PlaytestArena scoring is performed by an independent GPT-5.5 GUI judge~\citep{openai2026gpt55}. Given a generation prompt, a generated game, and its human-authored rubric, the judge loads the game in a browser, observes the rendered interface, and acts through clicks and keys. For each criterion, it issues a pass--fail judgment grounded in the gameplay trajectory and the expected behavior specified by the rubric. The final score is the fraction of criteria adjudicated as passed. This judge is disjoint from the Play2Code GUI playtester: the latter receives no rubric, produces only gameplay traces and actionable feedback for the Game Agent, and never determines the final benchmark score.

\paragraph{Is the GUI Evaluator Reliable?}
Because PlaytestArena scores are produced by an automated judge, we validate its decisions against blind human annotations over the full 200-game benchmark. At the criterion level, raw GPT-5.5--human agreement is 87.9\%, with $\kappa=0.66$; the corresponding human--human values are 91.4\% and $0.68$. At the game level, GUI- and human-derived scores are also strongly aligned (Spearman's $\rho=0.89$, Pearson's $r=0.89$), compared with $0.92$ and $0.93$ between humans. We further re-evaluate the same generated artifacts with Claude Opus 4.8~\citep{anthropic2026opus48} and Gemini 3.5 Flash~\citep{google2026gemini35flash}. All three judges preserve the ordering Direct LLM $<$ OpenGame $<$ Play2Code, and their game-level scores are strongly correlated. These results indicate that the main comparison is not an artifact of one judge's calibration. The full protocol, cross-judge scores, correlations, and failure analysis appear in Appendix~\ref{app:judge_validation}.

\begin{table}[h]
\centering
\small
\resizebox{\linewidth}{!}{
\begin{tabular}{lcc}
\toprule
\textbf{Metric} & \textbf{GUI--Human} & \textbf{Human--Human} \\
\midrule
Raw agreement (\%) & 87.9 & 91.4 \\
Cohen's $\kappa$ & 0.66 & 0.68 \\
Spearman's $\rho$ & 0.89 & 0.92 \\
Pearson's $r$     & 0.89 & 0.93 \\
\bottomrule
\end{tabular}
}
\caption{Agreement between the GPT-5.5 GUI judge and human evaluators over the full 200-game PlaytestArena benchmark.}
\label{tab:judge_agreement}
\vspace{-1em}
\end{table}


\section{Play2Code}
\label{sec:play2code}


\definecolor{methodA}{RGB}{240,240,240}  
\definecolor{methodB}{RGB}{220,235,247}  
\definecolor{methodC}{RGB}{253,237,200}  

\definecolor{rank1}{HTML}{FFF7D1}   
\definecolor{rank2}{RGB}{220,235,247}  
\definecolor{rank3}{RGB}{238,238,238}   

\setlength{\fboxsep}{2pt}  

\begin{table*}[h]
\centering
\small
\resizebox{\textwidth}{!}{%
\begin{tabular}{l ccc ccc ccc}
\toprule
& \multicolumn{3}{c}{\cellcolor{methodA}\textbf{Direct LLM}} & \multicolumn{3}{c}{\cellcolor{methodB}\textbf{OpenGame}} & \multicolumn{3}{c}{\cellcolor{methodC}\textbf{Play2Code (Ours)}} \\
\cmidrule(lr){2-4} \cmidrule(lr){5-7} \cmidrule(lr){8-10}
\textbf{Genre} & GPT-5.4 & Sonnet 4.6 & Kimi K2.5 & GPT-5.4 & Sonnet 4.6 & Kimi K2.5 & GPT-5.4 & Sonnet 4.6 & Kimi K2.5 \\
\midrule
Puzzle       & 35.2 & 33.4 & 31.8 & \colorbox{rank3}{60.5} & 59.7 & 50.2 & \colorbox{rank1}{77.8} & \colorbox{rank2}{74.6} & 52.8 \\
Strategy     & 30.4 & 28.6 & 26.9 & 54.3 & \colorbox{rank3}{55.1} & 44.1 & \colorbox{rank2}{69.5} & \colorbox{rank1}{70.2} & 52.0 \\
Card         & 36.8 & 34.7 & 32.5 & 61.2 & 60.6 & 51.4 & \colorbox{rank1}{76.8} & \colorbox{rank2}{75.9} & \colorbox{rank3}{67.3} \\
Action       & 28.3 & 26.5 & 24.8 & 51.7 & 51.2 & 41.3 & \colorbox{rank1}{69.4} & \colorbox{rank2}{67.8} & \colorbox{rank3}{58.6} \\
Platformer   & 31.5 & 29.8 & 27.4 & \colorbox{rank3}{55.6} & 55.1 & 45.2 & \colorbox{rank1}{74.7} & \colorbox{rank2}{70.9} & 52.4 \\
Management   & 33.6 & 31.5 & 29.7 & \colorbox{rank3}{58.4} & 57.9 & 48.5 & \colorbox{rank2}{73.2} & \colorbox{rank1}{73.7} & 54.5 \\
Shooter      & 27.4 & 25.6 & 23.9 & 50.8 & 50.3 & 40.5 & \colorbox{rank1}{67.9} & \colorbox{rank2}{66.8} & \colorbox{rank3}{57.4} \\
Other        & 29.5 & 27.7 & 25.6 & 53.2 & 52.6 & 42.7 & \colorbox{rank1}{69.3} & \colorbox{rank2}{68.5} & \colorbox{rank3}{59.8} \\
\midrule
\textbf{Avg.} & 31.6 & 29.7 & 27.8 & 55.7 & 55.3 & 45.5 & \colorbox{rank1}{72.3} & \colorbox{rank2}{71.1} & \colorbox{rank3}{56.9} \\
\bottomrule
\end{tabular}%
}
\caption{Per-genre rubric scores on PlaytestArena across three methods and three backbones, independently adjudicated by the GPT-5.5 GUI judge. Each cell reports the rubric pass-rate (\%). Cells shaded from \colorbox{rank1}{yellow} to \colorbox{rank3}{grey} mark the best, second-best, and third-best score per row.}
\label{tab:main_results_per_genre}
\vspace{-1em}
\end{table*}

PlaytestArena uses an independent GUI judge for final rubric scoring
(\S\ref{sec:playtestarena}). We now study a separate role for GUI agents:
producing a refinement signal inside the game-generation process.

To put the GUI agent in this role, we introduce \textbf{Play2Code}, a system that bridges game generation and playtesting and sustains their interaction across rounds (Figure~\ref{fig:play2code}). At its core, Play2Code couples two agents around a shared game runtime. A \emph{Game Agent} writes, debugs, and patches game code. A \emph{GUI playtester} plays the rendered build and reports what it experienced back to the Game Agent through shared memory. It does not receive the PlaytestArena rubric and does not produce the final benchmark score. Each generation task unfolds as a sequence of rounds, each consisting of a \emph{generate or revise} step by the Game Agent and a \emph{play-and-report} step by the GUI playtester.

The following section focuses on the conceptual description of the essential components in Play2Code. Further implementation details, including the workflows of the game agent and GUI agent and the management of the memory system, are provided in Appendix~\S\ref{app:system}.

\paragraph{Game Agent.} The game agent produces or revises code conditioned on the most recent build, the episode memory, and the cross-task skill and world memories. The output is written into a shared runtime, packaged as a self-contained HTML-based artifact that the GUI agent can immediately load.

\paragraph{GUI Agent.} The GUI agent loads the runtime in a browser and plays through it without access to the code or evaluation rubric, relying only on memory and a game guide. Once the GUI agent decides that the episode should end, it compiles its action and observation log into two artifacts. The first is a summary describing events and interaction behaviors during play. The second is a list of actionable fixes that map observed failures to concrete code-level changes, for example, enemies remain stationary when they should patrol $\rightarrow$ verify enemy patrol trigger conditions. Both are written into the shared subspace of the episode memory.

\paragraph{Between Rounds.} The game agent reads the summary and the fix list and decides, on its own, which fixes are worth acting on and which to defer or discard, treating the GUI agent's report as advice rather than instruction. A new round begins with the game agent revising the build accordingly. The loop terminates when the GUI agent judges that the build meets the intent of the prompt, or when a maximum number of rounds is reached.


\paragraph{Memory System.} Across many generation tasks, the same kinds of failures recur, such as incorrect jump arcs in platformers, yet each is often rediscovered from scratch. To preserve and reuse this experience, Play2Code organizes memory along two dimensions, \emph{lifecycle} (in-task or cross-task) and \emph{scope} (private, shared, or common), yielding three layers. \textbf{Episode Memory} is in-task, accumulating summaries, fixes, and attempts across rounds in a shared subspace between agents. \textbf{Skill Memory} is cross-task and agent-isolated, holding each agent's accumulated know-how, recurring code patterns for the game agent and interaction strategies for the GUI agent. \textbf{World Memory} is cross-task and shared across all agents, recording general game rules, common archetypes, and universal design principles. Together, these layers turn Play2Code into a system whose competence grows with experience.

\begin{figure*}[t]
    \centering
    \includegraphics[width=\linewidth]{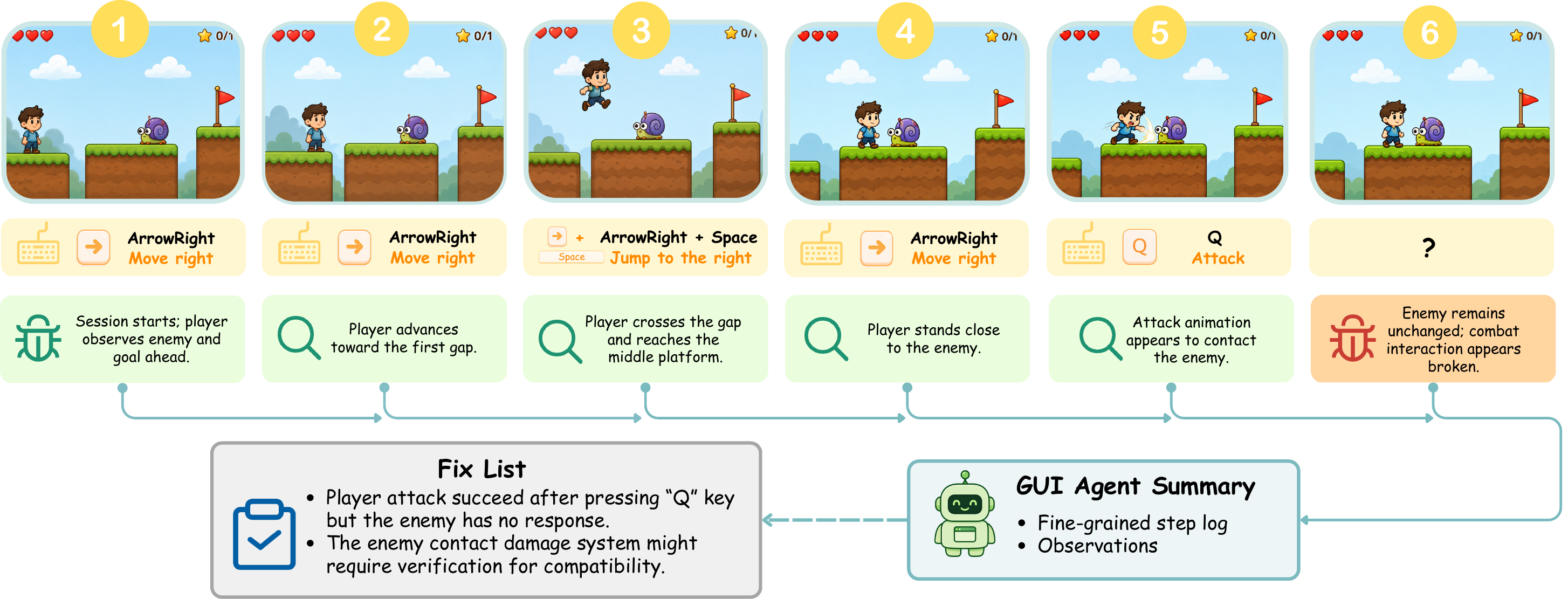}
   \caption{Example playtest trace produced by the GUI Agent. Per-step inputs and observations are distilled into a summary and an actionable fix list that the Game Agent consumes in the next revision round.}
    \label{fig:play_trace}
\end{figure*}

\section{Does Play Drive Code Generation?}
\label{sec: performance}
To test whether positioning a GUI agent inside the generation process matters, we compare Play2Code against two baselines representing prior paradigms for AI-driven game generation (Figure~\ref{fig:three_paradigms}). Direct LLM prompts the model to emit the entire game in a single pass. OpenGame runs an end-to-end agentic pipeline that iteratively builds, runs, and inspects the game at syntactic and compiling level. Play2Code adds the missing component: a GUI agent that plays each generated build and feeds prescriptive feedback back into the next round of generation. We instantiate each method with three backbone models used as both Game Agent and GUI playtester: Claude Sonnet 4.6, GPT-5.4, and Kimi K2.5. In all cases, final PlaytestArena scores are assigned separately by the independent GPT-5.5 GUI judge.



\paragraph{Playtesting exposes what code-level signals miss.}
Table~\ref{tab:main_results_per_genre} reports two comparisons that together isolate the value of playtesting. Across three backbones, Play2Code improves over Direct LLM from 29.7 to 66.8, reflecting the cost of committing to an artifact without ever observing play: any mismatch between intended and realized behavior goes uncorrected. The more revealing comparison is with OpenGame (52.2 → 66.8), an agentic system that already builds, runs, and inspects games iteratively, but does not play them. That Play2Code consistently outperforms OpenGame isolates the playtesting signal itself: failures such as unresponsive controls, missing state transitions, or win conditions that never trigger are invisible to code- and compilation-level inspection, and only surface during play. Direct ablations (Appendix~\ref{app:ablation}) confirm that both the GUI Agent and the layered memory system are essential components of Play2Code.

\begin{figure*}[t]
    \centering
    \begin{overpic}[width=\textwidth]{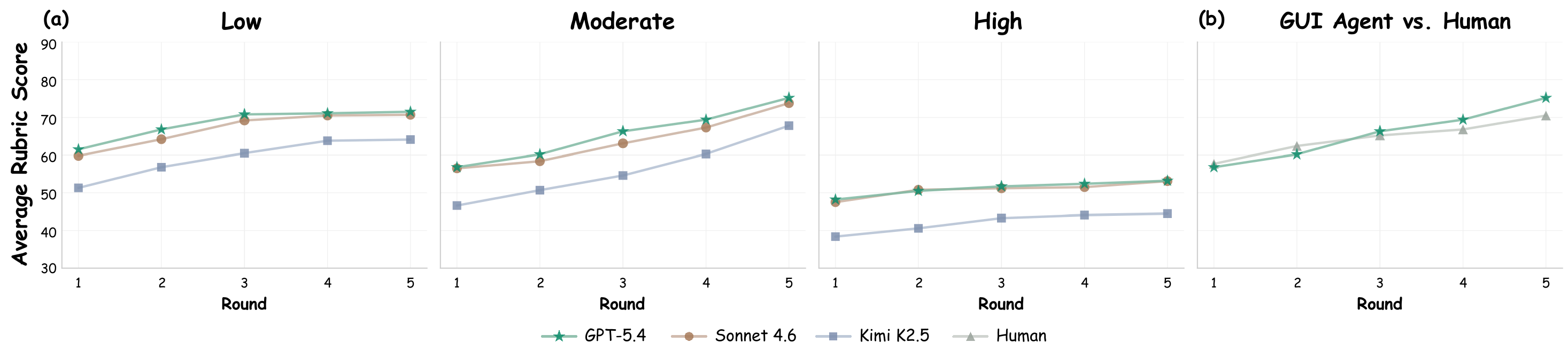}
      \put(10.2,20.1){\colorbox{white}{\scriptsize\bfseries Large gain}}
      \put(34.0,20.1){\colorbox{white}{\scriptsize\bfseries Moderate gain}}
      \put(59.2,20.1){\colorbox{white}{\scriptsize\bfseries Limited gain}}
    \end{overpic}
    \caption{Round-wise evolution of average rubric scores. (a) All 200 games are assigned to threshold-based empirical groups according to their observed five-round refinement gain: limited ($\Delta\leq10$), moderate ($10<\Delta\leq15$), or large ($\Delta>15$). These groups describe refinement responsiveness, not intrinsic game difficulty. (b) GUI-agent vs. human-driven refinement on the 50 moderate-gain games. See Appendix \S\ref{app:gui_vs_human} for details.}
    \label{fig:continual_playtesting}
    \vspace{-1em}
\end{figure*}

\paragraph{Continual playtesting drives round-wise game evolution.}
We further examine how Play2Code improves within a single evolving trajectory. As shown in Figure~\ref{fig:continual_playtesting}a, for each generated game, we record the rubric score after each play--code round. The scores increase monotonically from Round~1 to Round~3, showing that GUI feedback is actionable: failures observed are translated into concrete fixes that improve the next build. Early rounds often expose basic interaction failures, such as missing responses to player input, while later rounds address finer-grained issues like visual effects. This round-wise improvement suggests that playtesting provides a usable optimization signal for code generation. In this sense, Play2Code turns gameplay observations into an iterative refinement process.


\paragraph{Refinement effectiveness varies across tasks.}
Figure~\ref{fig:continual_playtesting}a reports all three empirical gain groups (see Appendix~\ref{app:gain_groups} for definitions). The limited-gain group improves by at most 10 points and plateaus early. The moderate-gain group improves steadily across the five rounds, while the large-gain group receives the greatest end-to-end benefit from iterative playtesting. Because the groups are defined by observed refinement gain, they should not be interpreted as an independent measure of game complexity or as evidence that complexity causally determines improvement. Instead, the analysis characterizes which tasks benefit more or less from the current Play2Code protocol.

\section{How Far Are GUI Agents from Human Playtesters?}
\label{sec:gui_vs_human}
Human playtesters remain irreplaceable for what they uniquely provide: a felt sense of difficulty, frustration, boredom, and surprise, alongside the tacit aesthetic judgments that no agent currently approximates. The question we ask here is therefore how far GUI agents are from filling that role: what they can deliver even when they fall short of a full human playtester, and where their behavior mirrors the idiosyncrasies of human playtesters themselves.


\paragraph{GUI agents produce more traceable feedback than humans.} A human playtester, even an attentive one, produces a report that is fundamentally a summary: salient moments remembered, causal chains reconstructed under uncertainty, and the raw trace of what actually happened, every key press, every observation that preceded a failure, lost by the time the report reaches the developer. The GUI agent's feedback is structured differently. Every action, observation, and diagnosis is logged (Figure~\ref{fig:play_trace}), so when the agent reports that ``enemies remain stationary when they should patrol,'' the failure is anchored in a recoverable sequence of frames rather than a playtester's reconstruction. 
This traceability translates into measurable gains: on games well-suited to GUI play, the agent approaches human playtesters in driving refinement (Figure~\ref{fig:continual_playtesting}b), suggesting that GUI-driven feedback can serve as a scalable and competitive refinement signal.

\begin{figure}[t]
    \centering
    \includegraphics[width=\linewidth]{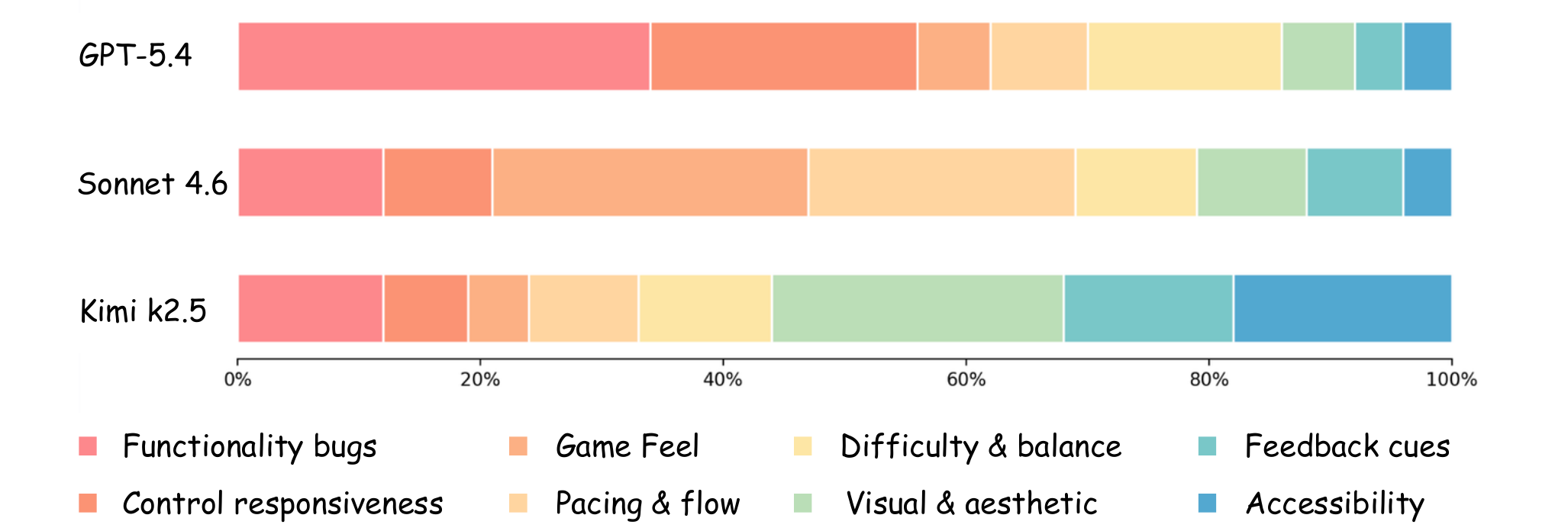}
    \caption{Distribution of feedback categories across playtested games for three backbones. 
    }
    \label{fig:feedback_distribution}
    \vspace{-1em}
\end{figure}

\paragraph{Different agents notice different things.}
Different GUI agents playing the same game produce feedback with different emphases. Figure~\ref{fig:feedback_distribution} shows this divergence across three playtester backbones during game generation. GPT-5.4 concentrates the bulk of its feedback on functionality bugs and control responsiveness, flagging broken mechanics, missing state transitions, and unresponsive controls. Sonnet 4.6 spreads its attention more evenly, devoting a comparable share to experiential issues such as pacing, game feel, and difficulty balance. Kimi K2.5 is the most striking outlier: nearly half its feedback concerns visual and aesthetic dimensions, where the others barely linger. Rather than establishing human-like taste, these results show that the feedback distribution depends substantially on the playtester backbone.

\section{Related Work}

\subsection{Game Generation and Evaluation}

Large language models have shown strong code-generation capabilities~\citep{hui2024qwen2,dong2025survey}, including synthesizing complete 
games~\citep{hu2024game,gallotta2024large}. Traditional code-generation evaluations often emphasize functional correctness~\citep{zheng2023codegeex}, which alone cannot capture whether a game is playable, responsive, or faithful to its intended interaction design~\citep{peng2026playcoder}. Recent work therefore evaluates and repairs GUI applications through interactive playthroughs~\citep{peng2026playcoder}. We adopt this interaction-centered view, with the GUI agent serving as a playtester whose observations drive subsequent revisions.

\subsection{GUI Agents}

GUI agents, typically built on vision-language models (VLMs), have demonstrated strong visual understanding and action planning~\citep{you2024ferret,qin2025ui,wang2025ui,huang2026learning}, interpreting screen-based observations and taking actions conditioned on the environment state and task goals~\citep{deng2023mind2web,wu2024copilot,xie2024osworld,liu2024autoglm,nguyen2025gui}. In game-related settings, multimodal and language-model agents have also been evaluated in dynamic environments and player simulations~\citep{ouyang2026gameworld,zhang2025beyond}.
These properties make GUI agents a natural candidate for 
scalable proxies of human feedback in real-time environments.

\subsection{Memory-Augmented Agent Evolution}

Agent evolution enables continual improvement through 
iterative feedback loops over execution, tool use, and 
multi-agent interaction in both software tasks and interactive environments~\citep{tao2024survey,gao2025survey,wang2025mobile,wang2025talk,wang2026co,huang2026ace,dong2026evaluation,wang2023voyager}.
In parallel, agent memory retains interaction histories, 
preferences, and task-specific knowledge for reuse across 
various 
tasks~\citep{park2023generative,shinn2023reflexion,zhong2024memorybank,chhikara2025mem0,yu2026agentic,pan2026m}, 
and can further support reusable skills that transfer 
successful strategies to new 
tasks~\citep{jiang2026xskill,ma2026skillclaw}.



\section{Conclusions}
We studied game generation under the constraint that quality lives in play, not in code. PlaytestArena scores generated games with an independent automated rubric evaluator that loads each build, plays it, and adjudicates human-authored criteria from observation. Play2Code uses a separate GUI playtester inside the generation loop: it receives no rubric and instead returns gameplay traces and actionable feedback to a Game Agent through shared memory. Across three backbones, Play2Code outperforms single-shot and agentic-coding baselines on PlaytestArena, with rubric scores rising monotonically within each refinement trajectory, evidence that playtesting yields not just a richer evaluation signal but also a usable optimization signal for code generation.

\section*{Limitations}

We note several limitations of this work.
First, our current implementation focuses on HTML-based games; extending the framework to broader game formats, such as native-engine or 3D titles, is left to future work.
Second, our benchmark covers a curated set of games selected for quality control, and does not exhaust the full diversity of real-world game scenarios.
Third, the memory schema is shared across all games in our current design, while game-specific or genre-aware memory structures may yield further gains.

\section*{Acknowledgments}
This work is supported by grants from the National Natural Science Foundation of China (No.~T260020055) and Computing for the Future at Fudan (CFFF).

\bibliography{custom}

\appendix

\clearpage
\newpage

\section*{Appendix}
\label{sec:appendix}

\section{GUI Playtester Feasibility Study: Details}
\label{app:gui_playtest_feasibility}

This appendix expands on the small-scale feasibility study 
summarized in \S\ref{sec:gui_playtest}, which establishes that 
GUI agents are operationally capable of playing games before 
we deploy them as playtesters in PlaytestArena 
(\S\ref{sec:playtestarena}) and as refinement drivers in 
Play2Code (\S\ref{sec:play2code}). We describe the testbed 
construction, level decomposition, agent and human protocols, 
and per-game results.

\subsection{Testbed Construction}
\label{app:feasibility_testbed}

The feasibility testbed contains 20 games, intentionally kept 
small so that the same levels can be played end-to-end by both 
GUI agents and human references under matched conditions. 
Games are drawn from two sources:

\begin{itemize}
    \item \textbf{LLM-assisted specifications (10 games):} Prompts authored by the same expert pool described in Appendix~\S\ref{app:playtestarena_construction}, but disjoint from the 200 PlaytestArena tasks. Each prompt is realized as a self-contained HTML/CSS/JS bundle by a frontier code model and lightly debugged by a human until it runs without crashes. These games are kept simple by design, since the purpose is to test whether agents can play, not whether they can play hard games.
    \item \textbf{Public web games (10 games):} Lightweight browser titles drawn from \texttt{itch.io}~\citep{itchio2026} and \texttt{Poki}~\citep{poki2026}, selected to span the five genres listed below. We use only games that run as static HTML bundles with no login, payment, or external API dependency.
\end{itemize}

The 20 games cover five genres: puzzle (5), strategy (4), 
card (4), platformer (4), and management (3). Compared to 
PlaytestArena, this testbed omits action and shooter genres 
because their fast-paced real-time dynamics introduce additional 
confounds (agent inference latency vs.\ game tick rate) that 
are orthogonal to the question of whether GUI agents can 
play at all.

\subsection{Level Decomposition}
\label{app:feasibility_levels}

Games vary substantially in length: a Sokoban-style puzzle may 
admit a clean per-puzzle decomposition, while a tower defense 
or management game runs as a single long session. To make 
\emph{pass} a meaningful unit of measurement, each game is 
decomposed into discrete \textbf{levels}, each with a single, 
unambiguous completion condition (e.g., ``reach the goal tile'', 
``clear all blocks'', ``survive wave 5'', ``reduce opponent HP to zero''). 
Decomposition is performed by the same expert who authored or 
selected the game, and reviewed by a second expert for 
unambiguity. Across the 20 games, this yields 118 levels 
in total (mean 5.9 per game, min 3, max 10), 
which we round to approximately 120 in the main text.

A level is counted as \textbf{passed} if its completion condition 
is satisfied within a single episode, where an episode terminates 
on (i) completion, (ii) an explicit game-over state, or 
(iii) a per-level timeout (5 minutes wall-clock for GUI agents, 
matched for humans). pass@$k$ is computed as the fraction of 
levels for which at least one of $k$ independent episodes 
satisfies this condition.

\subsection{Agent Protocol}
\label{app:feasibility_agent_protocol}

We connect three GUI agent backbones to each game: 
GPT-5.4, 
Claude Sonnet 4.6, and 
Kimi K2.5. Each backbone is run 
within the same GUI-agent scaffold used later by the GUI 
Playtester in Play2Code, 
with three differences appropriate to a feasibility study:

\begin{itemize}
    \item \textbf{No rubric, no fix list.} The agent is given only the game guide (controls, objective, success condition) and the rendered screen. It is not asked to produce a fix list or pass--fail rubric judgments; its sole task is to attempt the level.
    \item \textbf{No memory.} Each level is played from a clean slate, so pass@$k$ reflects raw playing capability rather than the benefit of accumulated experience.
    \item \textbf{Fixed observation rate.} Screenshots are captured at 1280$\times$720 resolution. The agent decides when to issue the next action; we do not impose a fixed tick.
\end{itemize}

For each (backbone, level) pair we run 20 independent episodes, 
allowing us to report pass@5, pass@10, and pass@20 from the 
same set of rollouts.

\subsection{Human Reference Protocol}
\label{app:feasibility_human_protocol}

We recruit three human players from the same pool as the 
expert annotators (Appendix~\S\ref{app:playtestarena_construction}), 
with no overlap with the prompt/rubric authors or the GUI judge 
validation annotators. Each human is given the same game guide 
provided to the GUI agents and plays each level under matched 
conditions: the same browser, the same 1280$\times$720 viewport, 
and the same 5-minute per-level timeout. Each human plays each 
level up to 20 times if they have not yet passed it, allowing 
a directly comparable pass@$k$ to be computed. Humans are 
compensated at the same hourly rate as the expert annotators. 

To prevent fatigue effects, no human plays more than 30 levels 
in a single session, and sessions are capped at 90 minutes 
with a mandatory 15-minute break.

\subsection{Per-Genre Results}
\label{app:feasibility_results}

Table~\ref{tab:feasibility_per_genre} breaks down pass@10 by 
genre. All three GUI backbones clear the majority of levels 
across every genre, with the gap to the human reference 
narrowest on puzzle and card games (which reward deliberation 
over reaction time) and widest on platformer games (where 
precise timing of jumps remains a bottleneck for current 
GUI agents). This pattern foreshadows the difficulty 
stratification later observed in Play2Code 
(Figure~\ref{fig:continual_playtesting}a): genres in which 
GUI agents play more reliably also benefit more from 
GUI-driven refinement.

\begin{table}[h]
\centering
\small
\resizebox{\linewidth}{!}{
\begin{tabular}{lcccc}
\toprule
\textbf{Genre} & \textbf{GPT-5.4} & \textbf{Sonnet 4.6} & \textbf{Kimi K2.5} & \textbf{Human} \\
\midrule
Puzzle      & 0.81 & 0.83 & 0.74 & 0.95 \\
Strategy    & 0.78 & 0.79 & 0.69 & 0.94 \\
Card        & 0.84 & 0.82 & 0.76 & 0.96 \\
Platformer  & 0.64 & 0.66 & 0.55 & 0.87 \\
Management  & 0.72 & 0.71 & 0.62 & 0.89 \\
\midrule
\textbf{Avg.} & 0.75 & 0.76 & 0.68 & 0.92 \\
\bottomrule
\end{tabular}
}
\caption{Pass@10 by genre on the 20-game feasibility testbed. All three GUI agents clear the majority of levels across every genre, with the largest gap to humans on platformer games.}
\label{tab:feasibility_per_genre}
\end{table}

\subsection{Qualitative Observations}
\label{app:feasibility_qualitative}

Beyond aggregate pass rates, we observe three recurring patterns in agent rollouts that are relevant to the playtester role assumed in later sections:

\begin{itemize}
    \item \textbf{Observation grounded in the rendered surface.} Across backbones, the per-step observation logs (e.g., ``snake head is just below the apple'', Figure~\ref{fig:gui_play}) consistently reflect what is on screen rather than what the agent expects to be on screen, including in failure states such as ``no apple visible''. This is a prerequisite for using the same observation logs as evidence for rubric adjudication in \S\ref{sec:playtestarena}.
    \item \textbf{Retry behavior under failure.} When an action does not produce the expected effect (e.g., a button click with no visible response), agents typically retry once or twice before exploring an alternative interaction, rather than either giving up immediately or repeating the same action indefinitely. This behavior is what makes the GUI agent's report distinguish between ``mechanic missing'' and ``mechanic present but hard to trigger'', a distinction that matters for fix-list generation in Play2Code.
    \item \textbf{Failure modes that explain the gap to humans.} The dominant failure modes are (i) imprecise timing on real-time games (especially platformers) and (ii) misreading small or low-contrast UI elements. They are mainly backbone-level limitations caused by inference latency.
\end{itemize}

These observations are consistent with the three capabilities 
identified in \S\ref{sec:gui_playtest} (\emph{observe}, 
\emph{reason}, \emph{act}) and support the use of GUI agents 
as a playtester surrogate in the rest of the paper.

\section{PlaytestArena Construction Details}
\label{app:playtestarena_construction}

This section documents the human-expert workforce 
behind PlaytestArena, summarized in §3, including 
recruitment, qualifications, the annotation protocol for 
prompts and rubrics, and quality control.

\subsection{Expert Recruitment and Qualifications}

We recruit five human experts to author the 200 
generation prompts and the paired rubrics. Experts were 
recruited from a pool of graduate students and recent 
graduates in computer science, human--computer 
interaction, and game design at our institution. To 
qualify, each expert was required to meet at least two 
of the following three criteria:

\begin{itemize}
    \item self-reported $\geq$10 hours per week of digital 
    game play over the past year;
    \item prior coursework in game design, HCI, or 
    interactive systems;
    \item prior experience implementing a playable game 
    (course project, game jam entry, or hobby project) 
    in any engine or framework.
\end{itemize}

The final five experts comprise three graduate students 
and two recent graduates; three have shipped at least 
one playable game prototype, and all five report regular 
play across multiple genres. None of the experts are 
authors of this paper, and none participated in the GUI 
judge validation study (Appendix~\ref{app:judge_validation}), 
ensuring those two annotation efforts are independent. 

\subsection{Genre and Topic Coverage}

The 200 game topics are distributed across the eight 
genres reported in Figure~\ref{fig:playtestarena_stats}. 
Within each genre, topics are seeded from two sources: 
(i) representative subgenres listed in 
\citet{wiki2026videogamegenres}, and (ii) titles 
browsable on the public web-game platforms itch.io and 
Poki~\citep{itchio2026,poki2026}. We use these sources 
only to enumerate genre conventions and topic 
archetypes; no game code, asset, or rubric is copied 
from existing titles. Topics are reviewed by the lead 
authors before being assigned to experts to remove 
duplicates and ensure each topic is implementable as a 
self-contained HTML/CSS/JS bundle within a single 
generation pass by a frontier model.

\subsection{Prompt Construction Protocol}

Each topic is assigned to one of the five experts as 
primary author. The expert writes a generation prompt 
that specifies, at minimum:

\begin{itemize}
    \item the stateful gameplay loop;
    \item win and loss conditions;
    \item the set of valid player inputs (keys, mouse 
    actions);
    \item expected feedback behavior on input (visual, 
    auditory, or state-change cues);
    \item any progression structure (levels, waves, 
    rounds).
\end{itemize}

Each prompt is then reviewed by a second expert who 
flags ambiguity, missing constraints, or 
non-self-contained references (e.g., "make it feel like 
\emph{[existing game]}" without further detail). Prompts 
flagged in this second pass are revised by the primary 
author or removed. The final 200 prompts are the result 
of an initial pool of 247 candidate prompts, of which 47 
were discarded.

\subsection{Rubric Construction Protocol}

Once a prompt is finalized, the same primary expert 
drafts a paired rubric. Each rubric criterion is written 
to state a single expected in-play behavior in concrete, 
observable terms (e.g., ``the game loads and renders 
without errors'', ``the spawned enemies move toward the 
player's position''). The expert is instructed to cover 
five dimensions: \emph{mechanics}, \emph{controls}, 
\emph{progression}, \emph{interface}, and 
\emph{visual feedback}, with a target of 6--10 criteria 
per game. The final rubric corpus contains 1{,}548 
criteria across the 200 games, with a mean of 7.7 
criteria per game (min 5, max 17).

Each rubric is then reviewed by a second expert against 
three checks:

\begin{itemize}
    \item \textbf{Concreteness}: every criterion describes 
    a single observable behavior, not a quality judgment 
    (e.g., ``the spawn rate increases each wave'' is 
    accepted; ``the difficulty curve feels right'' is 
    not);
    \item \textbf{Observability}: every criterion can in 
    principle be adjudicated by a tester who only sees 
    the rendered game, without access to source code or 
    internal state;
    \item \textbf{Faithfulness}: every criterion follows 
    from the paired prompt and is not introduced from the 
    expert's personal preferences.
\end{itemize}

Items failing any check are revised or removed. A final 
pass by one of the lead authors verifies all 200 
prompt--rubric pairs against the same three checks.

\subsection{Calibration and Inter-Expert Consistency}

Before authoring the main set, all five experts complete 
a calibration round on three shared topics (one puzzle, 
one strategy, one platformer). Each expert independently 
writes a prompt and rubric for these three topics; the 
five drafts are then reviewed jointly, and disagreements 
on what counts as a concrete, observable criterion are 
resolved in discussion. The calibration prompts and 
rubrics are discarded and are not part of the released 
benchmark.

As a post-hoc check on consistency, we have each expert 
re-review a random 10\% sample of criteria authored by 
their peers and mark each as \textsc{accept} or 
\textsc{revise}. The acceptance rate is 95.3\%, with the 
most common reason for \textsc{revise} flags being a 
criterion that is concrete but underspecified (e.g., 
omitting which input triggers a behavior). Flagged 
criteria were revised before benchmark release.

\section{GUI Evaluator Validation: Full Setup and Analysis}
\label{app:judge_validation}

This appendix expands the validation summarized in
\S\ref{sec:playtestarena}. We evaluate agreement with humans over all
200 PlaytestArena games, test whether method rankings are stable under
alternative GUI judges, and audit whether low scores arise from broken
games or from failures of the GUI agent to understand the interaction.

\subsection{Full-Benchmark Human Validation}

\paragraph{Annotators and protocol.}
We recruit three annotators in computer science with prior game-playing
experience (self-reported $\geq 5$ hours per week). None is an author of
this paper or participated in PlaytestArena construction. The validation
covers all 200 benchmark games rather than a sampled subset. Annotators
are blind to the generation method, the GPT-5.5 judge's verdicts, and its
play trace. For each game, they receive the generation prompt and full
human-authored rubric, play the build in a browser under the same
10-minute session budget as the GUI judge, and mark every criterion as
\textsc{pass} or \textsc{fail}. Before the study, annotators complete a
calibration round on three held-out games; these calibration data are
discarded.

\paragraph{Metrics.}
At the criterion level, we report raw agreement and Cohen's $\kappa$.
At the game level, each evaluator's score is the fraction of rubric
criteria marked \textsc{pass}; we report Pearson's $r$ over these scores
and Spearman's $\rho$ over the resulting rankings. Table~\ref{tab:full_human_validation}
shows that GPT-5.5--human agreement is close to human--human agreement
under all four metrics.

\begin{table}[t]
\centering
\small
\resizebox{\columnwidth}{!}{%
\begin{tabular}{lcc}
\toprule
\textbf{Metric} & \textbf{GPT-5.5--Human} & \textbf{Human--Human} \\
\midrule
Raw agreement (\%) & 87.9 & 91.4 \\
Cohen's $\kappa$ & 0.66 & 0.68 \\
Pearson's $r$ & 0.89 & 0.93 \\
Spearman's $\rho$ & 0.89 & 0.92 \\
\bottomrule
\end{tabular}
}
\caption{Agreement between the GPT-5.5 GUI judge and human annotators over the full 200-game benchmark.}
\label{tab:full_human_validation}
\end{table}

\subsection{Robustness to the Choice of GUI Judge}

We re-evaluate the same generated artifacts, while holding the prompts,
human-authored rubrics, and interaction protocol fixed, using Claude
Opus 4.8 and Gemini 3.5 Flash as alternative GUI judges. As shown in
Table~\ref{tab:cross_judge_scores}, the absolute scores vary modestly,
but all three judges preserve the same method ordering: Direct LLM
$<$ OpenGame $<$ Play2Code.

\begin{table}[t]
\centering
\small
\resizebox{\columnwidth}{!}{%
\begin{tabular}{lccc}
\toprule
\textbf{Method} & \textbf{GPT-5.5} & \textbf{Opus 4.8} & \textbf{Gemini 3.5 Flash} \\
\midrule
Direct LLM & 31.6 & 33.9 & 35.1 \\
OpenGame & 55.7 & 54.2 & 58.6 \\
Play2Code & \textbf{72.3} & \textbf{70.8} & \textbf{75.4} \\
\bottomrule
\end{tabular}%
}
\caption{Rubric pass-rates (\%) for the same GPT-5.4-generated artifacts under three independent GUI judges.}
\label{tab:cross_judge_scores}
\end{table}

Game-level scores are also highly correlated across judges
(Table~\ref{tab:cross_judge_corr}). Thus, the advantage of Play2Code is
not specific to the capabilities or calibration of GPT-5.5.

\begin{table}[t]
\centering
\small
\resizebox{\columnwidth}{!}{%
\begin{tabular}{lcc}
\toprule
\textbf{Judge Pair} & \textbf{Pearson's $r$} & \textbf{Spearman's $\rho$} \\
\midrule
GPT-5.5 vs. Opus 4.8 & 0.92 & 0.94 \\
GPT-5.5 vs. Gemini 3.5 Flash & 0.89 & 0.91 \\
Opus 4.8 vs. Gemini 3.5 Flash & 0.86 & 0.89 \\
\bottomrule
\end{tabular}%
}
\caption{Game-level score correlations between GUI judges.}
\label{tab:cross_judge_corr}
\end{table}

\subsection{Game Failure or GUI-Agent Failure?}

To separate artifact failures from evaluator limitations, we manually
inspect 50 GUI-agent gameplay trajectories spanning genres and
refinement-gain groups. We define a \emph{successful playthrough} as a
trajectory in which the agent (i) correctly identifies the game
objective, (ii) performs meaningful interactions consistent with the
intended mechanics, and (iii) reaches valid game states rather than
failing because it misunderstood how to play. Under this definition,
46 of 50 trajectories (92\%) are successful playthroughs. This audit
suggests that most low rubric scores reflect failures in the generated
artifacts, while also exposing cases in which agent capability remains a
confound.

\paragraph{Representative cases.}
The inspected trajectories exhibit three recurring patterns. First, in
a genuine artifact failure, the agent follows the stated controls and
reaches the relevant game state, but an expected behavior---for example,
enemy movement or a win-state transition---never occurs despite repeated
valid attempts. Second, in a clear agent failure, a small or low-contrast
control is visible and usable by a human, but the GUI agent overlooks it
and incorrectly concludes that the game cannot start. Third, some cases
are intrinsically ambiguous: in a timing-sensitive platformer, the agent
understands the objective and makes meaningful jump attempts, yet cannot
determine whether repeated failure is caused by an overly narrow timing
window or by its own action latency. We retain such cases as evaluator
limitations rather than treating every low score as definitive evidence
of a broken game.

\section{Play2Code System Design Detail}
\label{app:system}

This section documents the concrete instantiation of the Game Agent and
the GUI Agent introduced in \S4. The two agents share a runtime and a
memory system, but operate over disjoint observation spaces and tool
inventories, reflecting their respective roles as developer and
playtester.

\subsection{Game Agent}
\label{app:system:game_agent}

The Game Agent is responsible for translating the generation prompt into a
runnable HTML-based game artifact and for revising that artifact between
rounds based on the GUI Agent's report. It operates in a code-first mode,
working autonomously through a fixed multi-phase workflow rather than
producing the game in a single shot.

\paragraph{Inputs.} At the start of each round, the Game Agent receives:
(i) the original generation prompt, (ii) the current state of the
codebase (or an empty scaffold in the first round), (iii) the GUI Agent's
play summary and actionable fix list from the previous round (rounds
$\geq 2$ only), and (iv) the relevant slices of the shared memory system
(\S\ref{app:system:memory}).

\paragraph{Workflow.} In the first round, the Game Agent proceeds through five phases after receiving the user prompt: game design, asset generation, code implementation, build-time verification, and memory capture. In subsequent rounds, once the GUI Agent provides a play summary and actionable fix list, the Game Agent skips the game design and asset generation phases by default. Instead, it reviews the GUI feedback, decides which suggested fixes are relevant and feasible, and resumes from code implementation, followed by build-time verification and memory capture.

\begin{enumerate}
    \item \textbf{Game design.} A specific game design document is generated from the user requirement. This phase also includes game guide (\texttt{GAME\_GUIDE.md}) generation, which is later consumed by the GUI Agent.
    (\S\ref{app:system:gui_agent}).
    \item \textbf{Asset generation.} Sprites, tilemaps, and audio are generated according to the game design.
    \item \textbf{Code implementation.} The agent implements scenes, entities, and game logic according to the game design document.
    \item \textbf{Verification.} The agent runs \texttt{npm run build} and a headless test pass, fixing any TypeScript or runtime errors before handing the build to the GUI Agent.
    \item \textbf{Memory capture.} On completion, the agent writes any non-obvious fix patterns, pitfalls, or design decisions to skill or world memory for use in subsequent tasks.
\end{enumerate}

\paragraph{Tools.} The Game Agent has access to file-system tools
(\texttt{read\_file}, \texttt{write\_file}, \texttt{run\_shell\_command}),
project-specific code generation tools (\texttt{generate\_game\_guide}
and \texttt{generate-game-assets}), and the memory interface
(\texttt{memory\_query}, \texttt{memory\_save}).

\subsection{GUI Agent}
\label{app:system:gui_agent}

The GUI Agent is responsible for playing each candidate build and
producing structured feedback that the Game Agent consumes in the next
round. It does not read source code or inspect internal game state; its
judgment is grounded entirely in the rendered game surface. It also has
no access to the PlaytestArena rubric and never assigns the final
benchmark score.

\paragraph{Inputs.} At the start of each playtest session, the GUI Agent
receives: (i) screenshots of the current game frame, captured at
$1280{\times}720$ resolution; (ii) the \texttt{GAME\_GUIDE.md} produced
by the Game Agent, which specifies the intended controls, objectives,
mechanics, and success conditions; (iii) a fixed system prompt defining
its role as a playtester, its interaction conventions, and its stopping
criteria; and (iv) the relevant GUI-side skill memory and shared world
memory (\S\ref{app:system:memory}).

\paragraph{Workflow.} The GUI Agent proceeds through five phases:
initial observation, game start, interactive playtesting, assessment,
and memory capture. Like the Game Agent, it follows an explicit
step-wise workflow rather than producing a judgment from a single
screenshot.

\begin{enumerate}
    \item \textbf{Initial observation.} The agent observes the rendered
    game screen and uses short waits when necessary to distinguish
    loading screens from frozen or unresponsive builds.

    \item \textbf{Game start.} The agent starts the game by clicking a
    visible start affordance, pressing the instructed key, or proceeding
    directly if the game auto-starts.

    \item \textbf{Interactive playtesting.} The agent plays toward the
    stated objective using only mouse and keyboard actions. It verifies
    the game guide through interaction rather than assuming it is
    correct. For movement-heavy games, it holds keys long enough to
    produce visible displacement and combines keys when chorded actions
    such as run-and-jump are required. The agent does not stop at the
    first failure state; for platformer-style games, it must attempt at
    least two materially different retries before declaring the level
    blocked.

    \item \textbf{Assessment.} The agent summarizes whether the run was
    completed, reached an ending, blocked by a bug, or could not start.
    It evaluates the start flow, controls, core mechanics, progression,
    and ending, and records severity-tagged findings grounded in the
    observed trajectory.

    \item \textbf{Memory capture.} At the end of the session, the agent
    writes reusable interaction patterns, false-positive cases, and
    game-archetype-specific testing heuristics to memory for future
    playtests.
\end{enumerate}

\paragraph{Tools.} The GUI Agent has access to browser-level interaction
tools: \texttt{browser\_screenshot}, \texttt{browser\_click},
\texttt{browser\_drag}, \texttt{browser\_key}, \texttt{browser\_type},
\texttt{browser\_scroll}, and \texttt{browser\_wait}, as well as the
memory interface \texttt{memory\_save}. These tools mirror the input
channels available to a human playtester. The agent has no access to the
DOM, source code, internal variables, or console logs, ensuring that its
feedback reflects what playing the game is actually like rather than
what code inspection reveals.

\paragraph{Output.} At the end of each session, the GUI Agent emits a
single Markdown report with a fixed structure: run outcome and
confidence, probe signals considered, chronological interaction log,
per-dimension gameplay assessment, severity-tagged findings, the most
blocking issue, and a recommended fix direction. This report is parsed
downstream into the play summary and actionable fix list provided to the
Game Agent in the next round.

\subsection{Independent Benchmark Judge}

After a method finishes generating an artifact, a separate GPT-5.5 GUI
judge performs PlaytestArena evaluation. Unlike the Play2Code GUI
playtester, this judge receives the human-authored rubric and returns a
pass--fail verdict for every criterion. It does not participate in the
refinement loop, share Play2Code memory, or provide feedback to the Game
Agent. Thus, the agents that generate and refine a game are disjoint from
the agent that determines its reported benchmark score.

\subsection{Memory System}
\label{app:system:memory}

Both agents read from and write to the shared memory system described in
\S4. Memory entries are tagged with \texttt{layer}
(\texttt{episode-shared}, \texttt{skill}, or \texttt{world}),
\texttt{owner} (\texttt{game-agent}, \texttt{gui-player}, or
\texttt{shared}), \texttt{kind} (\texttt{pitfall},
\texttt{fix\_pattern}, \texttt{decision},
\texttt{interaction\_pattern}, \texttt{false\_positive},
\texttt{observation}), and \texttt{archetype}. Agents query memory at
the start of each task scoped by archetype and role, and write new
entries selectively at the end of each session, retaining only insights
with cross-task reuse value rather than one-off task details.

\section{Experimental Details}
\label{app:exp_details}

All three methods compared in Table~\ref{tab:main_results_per_genre} (Direct LLM, OpenGame, and Play2Code)
are run on the same set of 200 PlaytestArena tasks, but their effective budgets differ in structure rather than in raw model calls.

\paragraph{Direct LLM.} The model emits the entire game in a single pass from the generation prompt, with no runtime inspection and no opportunity for revision. Each task therefore consists of exactly one round.

\paragraph{OpenGame.} We follow the released configuration of \citet{jiang2026opengame}, which iteratively builds, runs, and inspects the game at the syntactic and compilation level. The loop terminates once the game compiles cleanly and hangs normally—i.e., it enters a stable running state without crashing or exiting—or when the maximum of five inspection rounds is reached. We do not modify this stopping criterion to keep the baseline faithful to its original design.

\paragraph{Play2Code.} We set the maximum number of evolving rounds to
$R_{\max}{=}5$, matching OpenGame for fairness. However, the loop is not
required to run for all five rounds: at the end of each play-and-report
step, the GUI Agent decides whether the current build already meets the
intent of the generation prompt. Concretely, the GUI Agent is prompted to
emit an explicit \texttt{Run Outcome} field with values in
\{\texttt{completed}, \texttt{reached-ending}, \texttt{blocked-by-bug},
\texttt{could-not-start}\}, together with a confidence level. When the GUI
Agent reports \texttt{completed} or \texttt{reached-ending} with
\texttt{high} confidence, the loop terminates early and the current build is
taken as the final artifact; otherwise, the loop continues until $R_{\max}$
is reached. Across the 200 tasks, the average effective round count is
$3.24$ (median $3.05$, std $0.72$), with $93.5$\% of tasks
terminating before $R_{\max}$.

\paragraph{Inference cost.}
We estimate cost by rerunning a randomly sampled 10\% subset of
PlaytestArena with GPT-5.4 as the shared backbone. Table~\ref{tab:cost}
reports the average inference cost per game. Play2Code incurs additional
cost because it interleaves generation with browser-based playtesting,
while the comparison with OpenGame isolates the incremental cost of
adding gameplay-grounded feedback to an agentic coding pipeline.

\begin{table}[t]
\centering
\small
\begin{tabular}{lc}
\toprule
\textbf{Method} & \textbf{Approx. cost/game (USD)} \\
\midrule
Direct LLM & \$5 \\
OpenGame & \$12 \\
Play2Code & \$17 \\
\bottomrule
\end{tabular}
\caption{Average inference cost per game on a randomly sampled 10\% subset of PlaytestArena using GPT-5.4 as the shared backbone.}
\label{tab:cost}
\end{table}

\section{Ablation Studies of Play2Code}
\label{app:ablation}
To isolate where Play2Code's gains come from, we ablate both the layered memory system (Episode, Skill, World) and the role of the GUI Agent. All ablations are conducted on PlaytestArena using GPT-5.4 as the backbone for both the Game Agent and the GUI Agent, and GPT-5.5 as the judge, with the same settings as in the main text. We report the same rubric-pass score as in the main text.

\paragraph{Memory Ablations.}
Play2Code organizes memory along two dimensions, \emph{lifecycle}
(in-task vs.\ cross-task) and \emph{scope} (private, shared, or common),
yielding three layers: Episode Memory, Skill Memory, and World Memory (\S\ref{sec:play2code}).
To test whether each layer carries its own weight, we evaluate four variants:
\begin{itemize}
    \item \textbf{No Memory.} The Game Agent sees only the current prompt, the
    previous build, and the GUI Agent's report from the immediately preceding
    round. Nothing accumulates across rounds or across tasks.
    \item \textbf{Episode Only.} The shared in-task workspace is retained
    (round history, summaries, fix lists, prior attempts), but cross-task
    Skill and World memories are disabled.
    \item \textbf{Episode + Skill.} Each agent additionally carries its own
    cross-task Skill Memory (role-specific patterns and strategies), but the
    cross-agent World Memory is disabled.
    \item \textbf{Full (Episode + Skill + World).} The configuration used in
    the main experiments.
\end{itemize}
Results are reported in Table~\ref{tab:ablation_memory}. Removing memory entirely causes the largest drop, confirming that even in-task accumulation is non-trivially useful: without it, the Game Agent often re-introduces fixes
it has already tried and discarded in earlier rounds. Adding Skill Memory yields a further gain, most pronounced on genres with recurring implementation pitfalls (e.g., platformer jump arcs, tower-defense wave scheduling), where
the Game Agent benefits from patterns crystallized on earlier tasks. World Memory contributes a smaller but consistent improvement, with the clearest gains on genres underrepresented in the agent's prior experience,
suggesting that shared design-level knowledge transfers across agents and across tasks.

\begin{table}[t]
\centering
\small
\begin{tabular}{lc}
\toprule
\textbf{Variant} & \textbf{Rubric Score} \\
\midrule
No Memory                       & 64.1 \\
Episode Only                    & 69.3 \\
Episode + Skill                 & 71.8 \\
Full (Episode + Skill + World)  & \textbf{72.3} \\
\bottomrule
\end{tabular}
\caption{Ablation of the layered memory system on PlaytestArena.
Each layer contributes additively, with Episode Memory accounting for the
largest share and World Memory providing genre-transfer benefits.}
\label{tab:ablation_memory}
\end{table}

\paragraph{GUI Agent Ablation.}
Beyond memory, we ablate the GUI Agent itself to quantify its contribution
to the iterative refinement loop.
In the \textbf{No GUI Agent} variant, the GUI Agent is removed entirely:
the Game Agent receives no external playtest feedback and instead performs
self-verification by re-reading its own generated code and reasoning about
potential failures.
All other settings, including the Full memory configuration, remain unchanged.

Results are shown in Table~\ref{tab:ablation_gui}.
Removing the GUI Agent causes a substantial drop in rubric score,
falling below even the \emph{No Memory} baseline from Table~\ref{tab:ablation_memory}.
This underscores that the GUI Agent's role is not merely one of logging:
its grounded, execution-based feedback surfaces runtime failures and
interaction-layer bugs that the Game Agent systematically misses during
self-verification---particularly rendering glitches, input-handling errors,
and timing-sensitive mechanics that only manifest during actual play.
The gap is widest on action and physics-heavy genres, where
behavioral correctness cannot be inferred from static code inspection alone.

\begin{table}[t]
\centering
\small
\begin{tabular}{lc}
\toprule
\textbf{Variant} & \textbf{Rubric Score} \\
\midrule
No GUI Agent (self-verify)              & 58.7 \\
Full          & \textbf{72.3} \\
\bottomrule
\end{tabular}
\caption{Ablation of the GUI Agent on PlaytestArena (Full memory configuration).
Replacing the GUI Agent with Game Agent self-verification causes a larger
drop than removing any single memory layer, highlighting the necessity of
grounded, execution-based feedback.}
\label{tab:ablation_gui}
\end{table}

Taken together, the ablations support that each memory layer is independently useful, with Episode Memory carrying
the largest share, Skill Memory accumulating role-specific operational know-how across tasks, and World Memory transferring design-level knowledge across agents and genres.
Furthermore, the GUI Agent ablation demonstrates that grounded playtest feedback is the single most impactful component of Play2Code, contributing more than any individual memory layer.

\section{Refinement-Gain Grouping}
\label{app:gain_groups}
We group the 200 PlaytestArena games by their observed response to
Play2Code. For each game, $\Delta$ is the absolute rubric-score gain
across five rounds, averaged over GPT-5.4, Sonnet 4.6, and Kimi K2.5. We
then apply fixed thresholds: the \emph{limited-gain group} has
$\Delta\leq10$, the \emph{moderate-gain group} has
$10<\Delta\leq15$, and the \emph{large-gain group} has $\Delta>15$.
These are threshold-based empirical groups, not terciles, and therefore
need not contain equal numbers of games; the moderate-gain group contains
50 games. The grouping describes which games benefit more or less from
the observed refinement process. It is not an independent measure of
intrinsic game difficulty and does not support a causal claim that game
complexity determines refinement gain. Figure~\ref{fig:continual_playtesting}a
reports trajectories for all three groups.

\section{GUI-Agent vs.\ Human Refinement}
\label{app:gui_vs_human}

To situate GUI-agent-driven refinement relative to a human upper bound,
we recruit human annotators to replace the GUI Agent in the Play2Code loop
and compare the two on the 50 moderate-gain tasks from PlaytestArena.

\paragraph{Annotator recruitment and qualifications.}
We recruited 5 graduate students in computer science,
all with prior experience in software development and casual game playing.
Annotators were informed that the games were AI-generated but were not told
the specific experimental hypothesis being tested, nor were they shown the
rubric criteria, to avoid bias toward rubric-optimizing reports.
Each task was assigned to exactly one annotator, who completed all rounds
for that task independently.

\paragraph{Protocol.}
The human condition follows the Play2Code protocol exactly, with one
substitution: the GUI Agent is replaced by the human annotator.
At each round, the annotator plays the current build and submits a free-form
fix report in place of the GUI Agent's structured output.
Annotators were given the same generation prompt as the Game Agent and
instructed to report (i)~what they could and could not do,
(ii)~any bugs or unexpected behaviors encountered, and
(iii)~concrete suggestions for the next revision.
No explicit memory entries are written on the human side; each annotator's
observations exist only within the fix report for that round.
The Game Agent side is unchanged: Episode, Skill, and World memories are
retained and updated exactly as in the standard Play2Code loop, and the
same $R_{\max}{=}5$ round budget applies.

\paragraph{Results.}
As shown in Figure~\ref{fig:continual_playtesting}b, human-driven refinement
achieves higher rubric scores than GUI-agent-driven refinement across all
rounds, confirming that the GUI Agent has not yet closed the gap with human
play coverage. The difference is most pronounced in later rounds, where
human annotators continue to surface subtle interaction failures that the
GUI Agent tends to miss. Nevertheless, the GUI Agent tracks the human
trajectory closely in early rounds and requires no human effort, suggesting
that further improvements to GUI-agent capability could close the remaining
gap.
\end{document}